\documentclass[review,11pt]{elsarticle}

\usepackage{lineno,hyperref}
\hypersetup{pdfauthor={Zhiqiang Liao}, colorlinks=true, linkcolor=blue, anchorcolor=red,citecolor=blue, filecolor=red, urlcolor=red}

\journal{ }
\date{ }

\usepackage{amssymb,amsthm,amsmath}
\usepackage{soul}
\usepackage{multirow}
\usepackage[flushleft]{threeparttable}
\usepackage[english]{babel}
\usepackage{mathtools}
\usepackage{enumitem}
\usepackage{float, rotating}
\usepackage{caption, subcaption} 
\usepackage[margin=1.5cm]{geometry}
\usepackage{tabulary}
\usepackage{booktabs}
\usepackage[gen]{eurosym}

\usepackage{ds}
\usepackage{algorithm}
\usepackage{algpseudocode}
\usepackage{makecell}
\usepackage{etoolbox}

\selectfont

\newcolumntype{K}[1]{>{\centering\arraybackslash}p{#1}}

\floatname{algorithm}{Algorithm}

\newcommand{\bftab}{\fontseries{b}\selectfont}

\newtheorem{theorem}{Theorem}

\newdefinition{axiom}{Axiom}
\newdefinition{definition}{Definition}

\begin{document}
\captionsetup[figure]{labelfont={bf},labelformat={default},labelsep=period,name={Fig.}}
\captionsetup[table]{labelfont={bf},labelformat={default},labelsep=period,name={Table}}

\begin{frontmatter}


\title{Structured Lasso for convex nonparametric least squares: \\An application to Swedish electricity distribution networks}

\author[firstaddress]{Zhiqiang Liao\corref{CorrAuthor}}
\cortext[CorrAuthor]{Corresponding author.}
\ead{zhiqiangliao@bnbu.edu.cn}

\author[secondaddress]{Zhaonan Qu}
\ead{zq2236@columbia.edu}

\address[firstaddress]{Faculty of Business and Management, Beijing Normal-Hong Kong Baptist University, Zhuhai, China}
\address[secondaddress]{Data Science Institute, Columbia University, New York, USA}

\begin{abstract}

We propose a structured Lasso (SLasso) method for variable selection in convex nonparametric least squares (CNLS) problems. Although Lasso is a popular technique for standard regression problems, we find that it is often unable to select variables efficiently for CNLS. Exploiting the unique structure of subgradients in CNLS problems, we develop the SLasso method by combining the $\ell_1$-norm and the $\ell_{\infty}$-norm. We further introduce a relaxed version of SLasso to simultaneously achieve model sparsity and predictive performance, where we can control the two effects of SLasso--variable selection and model shrinkage--using separate tuning parameters. A Monte Carlo study is implemented to verify the superior finite sample performance of the proposed approaches. We also use real data from Swedish electricity distribution networks to illustrate the effects of the proposed variable selection techniques. Results from simulation and empirical studies confirm that the proposed SLasso and relaxed SLasso methods perform favorably, generally leading to sparser and more accurate predictive models, relative to the standard Lasso methods in the literature.
\end{abstract}


\end{frontmatter}

\section{Introduction}\label{sec: intro}
Convex nonparametric least squares (CNLS) is a classical nonparametric regression problem with known shape constraints, dating back to the seminal paper written by \citet{hildreth1954point}. Imposing shape constraints such as monotonicity, concavity, and convexity on the regression function is a natural way to limit the complexity of many statistical estimation problems. Such shape constraints also arise naturally in many settings that span a diverse array of management and economics applications. Examples include portfolio selection \citep{hannah2013multivariate}, inventory management \citep{curmei2023shape}, and housing price prediction \citep{liao2024convex}. Other applications include productivity analysis \citep{kuosmanen2012stochastic2} and economics \citep{kuosmanen2020conditional}, where the production function is assumed to be concave and monotonic. A key advantage of the CNLS estimator over more traditional nonparametric estimation methods is that no extra tuning parameters (e.g., the bandwidth parameter in kernel estimation) are required. 

An important recent development of nonparametric convex regression is the estimation of a sparse model with good data fidelity when high-dimensional data is available. Examples exist in a wide range of applications, such as the efficiency analysis of manufacturing \citep{yagi2020shape}, sustainable development evaluation \citep{dai2023variable}, and energy regulation \citep{duras2023using}. When the data contains irrelevant variables for the outcome of interest, incorrectly incorporating them may harm the regression models' interpretation and predictive power 
\citep{stone1980optimal}. Variable selection, a classical problem in statistics and optimization, is the process of choosing the relevant variables and screening out the irrelevant variables in sparse data. There is a rich body of literature studying variable selection in linear models (see, e.g., \citealp{tibshirani1996regression}, \citealp{zou2006the}, \citealp{bertsimas2016best}, \citealp{hastie2020best}, and the references therein). However, variable selection under the framework of CNLS is a notoriously difficult problem. \citet{lee2020lasso} and \citet{dai2023variable} develop the least absolute shrinkage and selection operator (Lasso) based regularization for CNLS estimators by applying $\ell_1$-norm constraints on the subgradients. They show that these methods could make the subgradients of irrelevant variables small, but are inefficient for choosing a sparse convex regression model. \citet{bertsimas2021sparse} propose the $\ell_0$-norm regularization for sparse convex regression, where they solve the best subset problem through a mixed integer optimization (MIO) formulation. However, the MIO problem takes much longer to certify its optimality than the standard Lasso \citep{hastie2020best}, and, more importantly, the best subset method may suffer from the problem of overfitting \citep{mazumder2023subset}.

These difficulties motivate the current study, where we develop a structured Lasso (SLasso) method in the context of CNLS. Previous Lasso methods for CNLS directly impose an $\ell_1$-norm penalty on the subgradients, but this straightforward extension of Lasso ignores the matrix structure of the subgradients, where each column corresponds to a \emph{single} variable. As a result, such methods remain ineffective in selecting a sparse convex regression model. SLasso is distinct and precisely addresses this challenge: as shown in Section \ref{subsec: slasso}, it merges the $\ell_1$-norm and $\ell_{\infty}$-norm into a mixed $\ell_1/\ell_{\infty}$-norm, which encourages entire columns of the subgradient matrix to be zero, thereby screening out irrelevant variables. Moreover, the optimization problem of SLasso is continuous and convex, making it more stable than other classic variable selection techniques, such as XGBoost \citep{chen2016xgboost} and best subset selection \citep{bertsimas2021sparse}. 

The performance of a model for high-dimensional data is often evaluated using two criteria: (a) model sparsity--a parsimonious model is preferred when the number of variables is large, and (b) predictive accuracy--the parsimonious model should achieve good out-of-sample performance by avoiding under- or over-fitting. Lasso type methods, including the SLasso, have two corresponding effects: variable selection and model shrinkage, which contribute to selecting relevant variables and reducing overfitting, respectively. Importantly, these two effects are usually regulated by a single tuning parameter. However, it is well-known that it can be difficult for variable selection techniques to \emph{simultaneously} produce a parsimonious model while retaining the best predictive power (see, e.g., \citealp{meinshausen2007relaxed} for Lasso in linear regression). Part of the reason is that a Lasso tuning parameter that accurately selects the relevant variables may lead to \emph{over-shrinkage} of their corresponding parameters, resulting in underfitting (see Figure \ref{fig: error}). A few recent attempts have been made to tackle this dilemma in CNLS. For example, \cite{bertsimas2021sparse} combined the $\ell_0$-norm and $\ell_2$-norm to jointly achieve the two criteria, but their approach resulted in the need to solve a nonconvex problem. In this paper, we address the over-shrinkage challenge by further developing a convex relaxed version of our proposed SLasso framework. It employs two separate tuning parameters for variable selection and model shrinkage, effectively decoupling the two processes, which further improves the  variable selection and predictive accuracy of SLasso. 

This study is also motivated by an application in the Swedish electricity distribution networks, where our primary goal is to estimate a cost function that is convex and non-decreasing. Swedish Energy Markets Inspectorate (SEMI) has identified over 40 potential input variables after interviews with several electricity distribution system operators (SEMI, \citeyear{swedish2021}). Estimating a sparse cost model based on such a large number of variables is immensely challenging due to the effects of correlation between the input variables. The need for a reliable variable selection method highlights the empirical relevance of this research.

We summarize the main contributions as follows.
\begin{enumerate}
    \item We propose the SLasso method which combines $\ell_1$-norm and $\ell_{\infty}$-norm penalties. It aims to eliminate entire columns from the subgradients matrix and thereby screening out the irrelevant variables. To determine the regularization parameters, we introduce adaptive weights to the $\ell_1/\ell_{\infty}$-norm penalty, leading to an estimator with a single tuning parameter but assigns different weights to distinct variables. We also derive convex optimization formulations for efficiently solving the SLasso problems (e.g., using commercial solvers like MOSEK and Gurobi). 
    \item To balance the effects of variable selection and model shrinkage, we propose relaxed SLasso, a two-stage procedure that extends the SLasso framework. We derive explicit optimization formulations to solve the relaxed SLasso problems, and investigate asymptotic behaviors of the relaxed SLasso. We also propose a data-driven parameter tuning procedure and develop a two-stage algorithm to select the parameters. By comparing the finite sample performance of different methods in Monte Carlo (MC) simulations, we demonstrate that the proposed relaxed SLasso achieves the best overall performance.
    \item This paper further contributes to the literature on variable selection for cost function estimation in empirical applications. In Sweden, SEMI acts as the energy regulator and determines the relevant production variables (e.g., delivered energy) and contextual variables (e.g., climate) that can be used in the energy regulation problem. However, there is severe multicollinearity between production variables (see the \citeyear{swedish2021} SEMI report). In this paper, we use data provided by SEMI to investigate the performance of various variable selection techniques. The empirical results show that the proposed relaxed SLasso achieves the best performance at estimating a sparse cost function while delivering superior predictive power, and these findings are consistent with our conclusions from the MC study.
\end{enumerate}

In Section \ref{sec: method}, we propose two new structured Lasso approaches,  SLasso and relaxed SLasso, and formulate the convex optimization problems for these methods. We also study the asymptotic properties of the relaxed SLasso. In Section \ref{sec: mc}, we evaluate the finite sample performance of the proposed methods through MC simulations. An empirical application is prepared in Section \ref{sec: app}. Section \ref{sec: concl} concludes this paper.

%

\section{Methodological framework}\label{sec: method}

\subsection{Convex nonparametric least squares}\label{subsec: cnls}
Suppose we observe $n$ pairs of independent and identically distributed (\emph{i.i.d.}) input and output data $\{(\boldsymbol{x}_i,y_i)\}_{i=1}^n\in \mathbb{R}^d\times\mathbb{R}$. We consider the following nonparametric convex regression model
\begin{equation} \label{regression}
    y_i = f_0(\boldsymbol{x}_i) + \varepsilon_i \quad \forall i = 1, \ldots, n,
\end{equation}
where $\varepsilon_i$ is a random variable with $E(\varepsilon_i)=0$ and $\text{Var}(\varepsilon_i)=\sigma^2 < \infty$. The goal is to estimate an unknown \emph{convex} function $f_0: \mathbb{R}^d\rightarrow\mathbb{R}$ with the given $n$ observations. Least squares is arguably the most natural way to fit the convex function $f_0$, and the convex nonparametric least squares (CNLS) estimator is defined as the solution to the following optimization problem
\begin{equation}\label{cnls}
    \min_{f\in\mathcal{F}} \frac{1}{2} \sum_{i=1}^n (y_i-f(\boldsymbol{x}_i))^2 ,
\end{equation}
where $\mathcal{F}$ is the space of globally convex functions on $\mathbb{R}^d$. In a high-dimensional setting, $d$ can be large when we have access to many covariates that are potentially relevant to the outcome variable. However, a reasonable assumption is that only a much smaller subset of covariates $S\subseteq\{1,\ldots,d\}$ with $|S|=s\leq d$ is important. In other words, $f_0$ is constant when varying covariates outside of $S$. We call $S$ the \emph{support} of $f_0$, and we aim to choose a subset of variables in order to recover it. 
In analogy to the linear regression model, a natural idea is to add a sparse penalty to problem \eqref{cnls}, leading to an objective of the form $\min_{f\in\mathcal{F}} \frac{1}{2} \sum_{i=1}^n (y_i-f(\boldsymbol{x}_i))^2 + \lambda P(f)$, where $P(\cdot)$ is a penalty function that encourages sparsity  and $\lambda\geq 0$ is the regularization (tuning) parameter.\footnote{One can recover the original CNLS problem as $\lambda$ approaches zero \citep{kuosmanen2010data}.} We refer readers to \cite{hastie2020best} for a more complete discussion of sparse penalty functions. In this paper, we leverage a reformulation of CNLS to impose sparsity on the support of $f_0$, which we explain next.

The infinite-dimensional optimization problem \eqref{cnls} can typically be reduced to a finite-dimensional convex problem characterized by a piecewise affine convex function with $n$ supporting hyperplanes (see, e.g. \citet{kuosmanen2008representation}, \citet{seijo2011nonparametric}, \citet{lim2012consistency}). More precisely, 
\eqref{cnls} can be reformulated as the following quadratic programming problem
\begin{equation}\label{cnls-quadratic}
\begin{aligned}\min_{\boldsymbol{\xi}_1,\ldots,\boldsymbol{\xi}_n;\, \boldsymbol{\theta}} \quad &  \frac{1}{2} \sum_{i=1}^n(y_i-\theta_i)^2 \\
    \mbox{\textit{s.t.}}\quad
    & \theta_i + \boldsymbol{\xi}_i^T(\boldsymbol{x}_j-\boldsymbol{x}_i)\leq \theta_j,  \quad i,j=1,\ldots,n, \qquad 
\end{aligned}  
\end{equation}
where $\theta_i$ represents the value of $f(\boldsymbol{x}_i)$, and $\boldsymbol{\xi}_i\in\mathbb{R}^d$ is the subgradient or coefficient of a convex function $f\in \mathcal{F}$ at point $\boldsymbol{x}_i$. The constraints in \eqref{cnls-quadratic} are the so-called shape constraints and impose convexity (or, equivalently, concavity by setting ``$\geq$'' in the constraints), as any convex function satisfies the conditions  
\[f(\boldsymbol{x}_i) + \partial f(\boldsymbol{x}_i)^T(\boldsymbol{x}_j-\boldsymbol{x}_i)\leq f(\boldsymbol{x}_j),  \quad i,j=1,\ldots,n.\]
Given the solution $\{\hat{\boldsymbol{\xi}}_i, \hat{\theta}_i\}_{i=1}^n$ to problem \eqref{cnls-quadratic}, we can express the convex function $\hat{f}\in \mathcal{F}$ that solves the CNLS problem \eqref{cnls} as \citep{kuosmanen2008representation}
\begin{equation}\label{estimator}
    \hat{f}(\boldsymbol{x})=\max_{i=1,\ldots,n} \big\{\hat{\theta}_i+ \hat{\boldsymbol{\xi}}_i^T(\boldsymbol{x}-\boldsymbol{x}_i)\big\}.
\end{equation}
Note that by setting $\boldsymbol{\xi}_i=\boldsymbol{\xi}_j,$ $\forall i, j = 1,\ldots,n$ in \eqref{cnls-quadratic}, we obtain the standard 
least squares regression. 

In this paper, we consider the variable selection problem for CNLS by leveraging the formulation \eqref{cnls-quadratic}. More precisely, we consider adding a penalty function $P$ of the gradient variables $\boldsymbol{\xi}_i$ in the following problem
\begin{equation}\label{pcnls}
\begin{aligned}\min_{\boldsymbol{\xi}_1,\ldots,\boldsymbol{\xi}_n;\, \boldsymbol{\theta}} \quad &  \frac{1}{2} \sum_{i=1}^n(y_i-\theta_i)^2 + \lambda P(\boldsymbol{\xi}_1,\ldots,\boldsymbol{\xi}_n) \\
    \mbox{\textit{s.t.}}\quad
    & \theta_i + \boldsymbol{\xi}_i^T(\boldsymbol{x}_j-\boldsymbol{x}_i)\leq \theta_j,  \quad i,j=1,\ldots,n. \qquad 
\end{aligned}  
\end{equation}
The objective function in problem \eqref{pcnls} takes the form of \textit{loss} + \textit{penalty}, and $\lambda$ is the tuning parameter that can adjust the amount of penalty or regularization on the subgradients. The most popular choice for the penalty $P$ is using Lasso, which adds an $\ell_1$-norm regularization \citep{tibshirani1996regression}, due to its computational and theoretical advantages compared with other variable selection methods like best subset and forward stepwise (\citealp{hastie2020best}). Adding this $\ell_1$-norm regularization, \cite{lee2020lasso} consider a Lasso type CNLS estimator as follows 
\begin{equation}\label{lasso}
\begin{aligned}\min_{\boldsymbol{\xi}_1,\ldots,\boldsymbol{\xi}_n;\, \boldsymbol{\theta}} \quad &  \frac{1}{2} \sum_{i=1}^n(y_i-\theta_i)^2 + \lambda \sum_{i=1}^n \|(\xi_i^1, \xi_i^2, \ldots, \xi_i^d)\|_{1} \\
    \mbox{\textit{s.t.}}\quad
    & \theta_i + \boldsymbol{\xi}_i^T(\boldsymbol{x}_j-\boldsymbol{x}_i)\leq \theta_j, \quad  i,j=1,\ldots,n, \qquad 
\end{aligned}  
\end{equation}
where $\|\cdot\|_1$ is the $\ell_1$-norm. For $i=1,\ldots,n$ and $k=1,\ldots,d$, let $\xi_i^k$ denote the $k$-th element of the vector $\boldsymbol{\xi}_i$. Note that there are $nd$ individual terms $|\xi_i^k|$, for $i=1,\ldots,n$ and $k=1,\ldots,d$, in the penalty function in \eqref{lasso}. 

Although \eqref{lasso} is a natural extension of Lasso to CNLS, it does not fully take into account the structure in the CNLS problem. To properly institute variable selection, we need to consider the matrix $\boldsymbol{\xi}=[\boldsymbol{\xi}_1,\ldots,\boldsymbol{\xi}_n] \in n\times d$ of subgradient variables in \eqref{cnls-quadratic}. Specifically, for $i=1,\ldots,n$ and $k=1,\ldots,d$, let
\begin{equation}\label{matrix}
    \boldsymbol{\xi} = (\xi_i^k)_{n\times d}=
    \begin{bmatrix}
    \xi_1^1 & \xi_1^2 & \ldots & \xi_1^d \\
    \xi_2^1 & \xi_2^2 & \ldots & \xi_2^d \\
      &   & \ddots &   \\
    \xi_n^1 & \xi_n^2 & \ldots & \xi_n^d \\
    \end{bmatrix}.
\end{equation}
Note that each column of $\boldsymbol{\xi}$ corresponds to the subgradients with respect to a distinct variable. As a result, sparsity in the CNLS model suggests a structured (overlapping) relationship between the $nd$ variables $\xi_i^k$: it requires a column (of $n$ individual entries) in matrix $\boldsymbol{\xi}$ to be \emph{simultaneously} selected or excluded. In other words, the selected nonzero elements of $\boldsymbol{\xi}$--also referred to as the supports of the subgradient of $f$--should share this overlapping structure. For example, assuming the cardinality of the support $S$ to be $s=d-1$, we would like to obtain a sparse model with the estimates as follows
\begin{equation}\label{matrix estimates}
    \hat{\boldsymbol{\xi}} = (\hat{\xi}_i^k)_{n\times d}=
    \begin{bmatrix}
    \hat{\xi}_1^1 & \hat{\xi}_1^2 & \ldots & \hat{\xi}_1^d \\
    \hat{\xi}_2^1 & \xi_2^2 & \ldots & \hat{\xi}_2^d \\
    \makebox(5,0){\rule[5ex]{0.8pt}{4\normalbaselineskip}}  &   & \ddots &   \\
    \hat{\xi}_n^1 & \hat{\xi}_n^2 & \ldots & \hat{\xi}_n^d \\
    \end{bmatrix},
\end{equation}
where an entire column of the matrix $\hat{\boldsymbol{\xi}}$ is excluded from the model, i.e., set to zero. Similarly, with higher levels of sparsity, other columns of the estimated subgradients could also be set to zero. In contrast,  the standard $\ell_1$-norm in \eqref{lasso} imposes sparsity on individual coefficients rather than enforcing the column-wise sparsity as expected in \eqref{matrix estimates}. As a result, it may not be able to exclude entire columns of $\boldsymbol{\xi}$, leading to inefficient variable selection in CNLS problems. In fact, our experiments show that adding the $\ell_1$-norm to CNLS regression can help make the subgradients of irrelevant variables \emph{small}, but it cannot reduce them to zero efficiently; see \cite{xu2016faithful} for a similar observation. Our structured Lasso approach precisely tackles this challenge.

\subsection{Structured Lasso}\label{subsec: slasso}

In this section, we propose the structured Lasso (SLasso) method for variable selection in CNLS problems. Let $(\xi_1^k, \xi_2^k, ..., \xi_n^k)$, $k = 1,...,d$ be the column-wise components in the matrix $\boldsymbol{\xi}$ defined in \eqref{matrix}. We define the $\ell_1/\ell_{q}$-norm that combines $\ell_1$-norm and $\ell_{q}$-norm as:
\begin{equation}\label{slasso}\|\boldsymbol{\xi}\|_{\ell_1/\ell_{q}} := \sum_{k=1}^d \|(\xi_1^k, \xi_2^k, \ldots, \xi_n^k)\|_{q},
\end{equation}
where $\|\cdot\|_{q}$ is the $\ell_{q}$-norm. A key ingredient in $\ell_1/\ell_q$-norm is the combination of two norms where $\|\cdot\|_{q}$ operates on all elements of the $k$-th column in matrix \eqref{matrix} and the sum of the $d$ $\ell_q$-norms imposes the $\ell_1$-norm penalty. When setting $q=1$ in \eqref{slasso}, the $\ell_1/\ell_q$-norm reduces to the Lasso penalty term used by \cite{lee2020lasso}, where it operates on the \textit{component-wise} subgradients and ignores their \textit{column-wise} sparsity structure. For $q=2$, the penalty term in \eqref{slasso} is often referred to as the ridge regularization in CNLS, which has been applied to the CNLS estimator to alleviate the overfitting problem \citep{keshvari2017penalized,bertsimas2021sparse}. 

This paper focuses on the case of the $\ell_1/ \ell_\infty$-norm, where we set $q$ as infinity in problem \eqref{slasso}. The main motivation for using the $\ell_1/ \ell_\infty$-norm penalty stems from the \textit{shared sparsity} among the columns of the subgradient matrix $\boldsymbol{\xi}$. Intuitively, the $\ell_\infty$-norm computes the maximum absolute value over the elements $(\xi_1^k, \xi_2^k, \ldots, \xi_n^k)$ for a given column $k=1,\ldots,d$. Then the $\ell_1$-norm penalty encourages sparsity over the maximum absolute values, which in turn translates to sparsity over entire columns of $\boldsymbol{\xi}$. As we shall see, imposing the column-wise sparsity by eliminating all subgradients for the corresponding irrelevant columns is the key behind the superior performance of the $\ell_1/ \ell_\infty$-norm for variable selection in CNLS regression.

We are now ready to formally define the structured Lasso (SLasso) for CNLS. For given tuning parameters $\lambda_k\geq0$ (to be determined), the SLasso method with the $\ell_1/ \ell_\infty$-norm penalty solves the following problem
\begin{equation}\label{cnls-slasso}
\begin{aligned}
    \min_{\boldsymbol{\xi}\in \mathbb{R}^{n\times d}; \boldsymbol{\theta}} \quad &  \frac{1}{2} \sum_{i=1}^n(y_i-\theta_i)^2 + \sum_{k=1}^d \lambda_k\|(\xi_1^k, \xi_2^k, \ldots, \xi_n^k)\|_{\infty} \\
    \mbox{\textit{s.t.}}\quad
    & \theta_i + \boldsymbol{\xi}_i^T(\boldsymbol{x}_j-\boldsymbol{x}_i)\leq \theta_j, \quad  i,j=1,\ldots,n. \qquad 
\end{aligned}  
\end{equation}
SLasso controls the complexity of CNLS models using multiple parameters $\lambda_k$ instead of one parameter $\lambda$ as is the case in the linear Lasso method. By using variable-dependent parameters $\lambda_k$, we can assign distinct weights to the variables, taking into account their relative magnitudes. Previously, \cite{liao2024convex} have considered the $\ell_1/ \ell_\infty$-norm penalty for variable selection in the context of convex regression. However, as their primary concern is the problem of overfitting, they focus on a convex version of the support vector regression (CSVR), with variable selection a secondary extension. In this work, instead of applying the $\ell_1/ \ell_\infty$-norm penalty to the ``$\epsilon$-insensitive'' objective in CSVR, we apply it directly to the squared loss objective in CNLS. Moreover, we also make use of \emph{adaptive} weights when selecting tuning parameters $\lambda_k$, which is crucial for ensuring that the penalty does not overly simplify the model by excluding potentially important variables (see \eqref{eq:adaptive-weights} below).

It is worth noting that the way SLasso encourages sparsity in CNLS models in \eqref{cnls-slasso} is reminiscent of, yet fundamentally distinct, from the group Lasso (see, e.g., \citealp{negahban2011simultaneous} and \citealp{chen2020on}). For a pre-defined set of groups of variables, the group Lasso imposes an $\ell_{1}$-regularization  over the $\ell_{2}$ norms of each group of coefficients. In contrast, the $\ell_1/\ell_{\infty}$-norm in SLasso applies penalization to all variables instead of implementing shrinkage on groups of variables. Still, SLasso does share conceptual similarities as the group Lasso, especially for the $\ell_1/\ell_{2}$-norm: we can understand each column of $\boldsymbol{\xi}$ as a distinct ``group'' of variables on which we impose sparsity. In this sense, the group structure in SLasso is pre-defined by the subgradient structure of the CNLS model, and naturally aligns with the $\ell_1/\ell_{\infty}$-norm (see our discussion in Section \ref{subsec: cnls}).

Computationally, the $\ell_1/\ell_{\infty}$-norm is continuous and convex, and therefore problem \eqref{cnls-slasso} remains a convex optimization problem. This is a non-trivial property, as discontinuities can result in instability in model prediction, and nonconvexity can increase the computational complexity. Within the SLasso framework, we can use modern convex optimization techniques to solve \eqref{cnls-slasso} at low computational costs, such as interior point solvers implemented in MOSEK and Gurobi, as discussed in \citet{seijo2011nonparametric}. 

To determine the regularization parameters $\lambda_k$ in practice, we apply adaptive weights. Let $\widetilde{\boldsymbol{\xi}}$ be the standard CNLS estimates from \eqref{cnls-quadratic}. Note that $\widetilde{\boldsymbol{\xi}}$ is well-defined  under general regularity conditions. Define the data-dependent weights $\boldsymbol{w}=(w_1,w_2,\ldots,w_d)\in \mathbb{R}^d$ for SLasso as
\begin{equation}
\label{eq:adaptive-weights}
    w_k := 1/\|(\widetilde{\xi_1^k}, \widetilde{\xi_2^k}, \ldots, \widetilde{\xi_n^k})\|_2, \quad k=1,\ldots,d.
\end{equation}
We then incorporate this prior weight information to penalize different subgradients in the SLasso, solving
\begin{equation}\label{aslasso}
\begin{aligned}
    \min_{\boldsymbol{\xi}\in \mathbb{R}^{n\times d}; \boldsymbol{\theta}} \quad &  \frac{1}{2} \sum_{i=1}^n(y_i-\theta_i)^2 + \lambda \sum_{k=1}^d w_k \|(\xi_1^k, \xi_2^k, \ldots, \xi_n^k)\|_{\infty} \\
    \mbox{\textit{s.t.}}\quad
    & \theta_i + \boldsymbol{\xi}_i^T(\boldsymbol{x}_j-\boldsymbol{x}_i)\leq \theta_j, \quad  i,j=1,\ldots,n, \qquad 
\end{aligned}  
\end{equation}
where $\lambda\geq0$ is a tuning parameter and $\lambda_k := \lambda{w_k}$ in \eqref{cnls-slasso}. To determine $\lambda$, we use cross-validation on a grid $\lambda_1< \lambda_2 < \dots<0$ with $\lambda_1 = \|\boldsymbol{X}^T\boldsymbol{y}\|_{\ell_1/\ell_{\infty}}$. Let $\hat{f}$ be an estimator of $f$ and consider the loss function
\begin{equation*}
    \textit{loss}(\lambda): = \frac{1}{m} \sum_{i=1}^m({\hat{f}(x_i)-y_i})^2
\end{equation*}
based on $m$ samples. In $k$-fold cross-validation, we randomly split the $n$ samples into $k$ subsets of size around $m=n/k$. For each subset, we compute the loss function using samples from that subset and $\hat f$ constructed by solving \eqref{aslasso} based on samples from  the rest of the subsets (\citealp{hastie2009elements}). We then choose the optimal $\lambda$ by minimizing the average out-of-sample $\textit{loss}(\cdot)$ across all $k$ folds.

Problem \eqref{aslasso} continuously shrinks entire columns of the subgradient matrix $\boldsymbol{\xi}$ toward zero as $\lambda$ increases, and some subgradients are exactly shrunk to zero when $\lambda$ is sufficiently large, thereby achieving model sparsity. The use of adaptive weights $w_k$ is crucial in our framework to address over-penalization of the model that incorrectly exluces relevant varibles. More precisely, if a feature is relevant, but the maximum magnitude of its subgradients is relatively small, it may be excluded if $\ell_{1/\infty}$-norm penalty is applied naively. By using a feature-specific weight $w_k$ that adapts to the magnitude of the subgradients (in a standard CNLS regression), we effectively account for these variations across features.

\begin{figure}[H]
    \centering
    \includegraphics[width=0.5\linewidth]{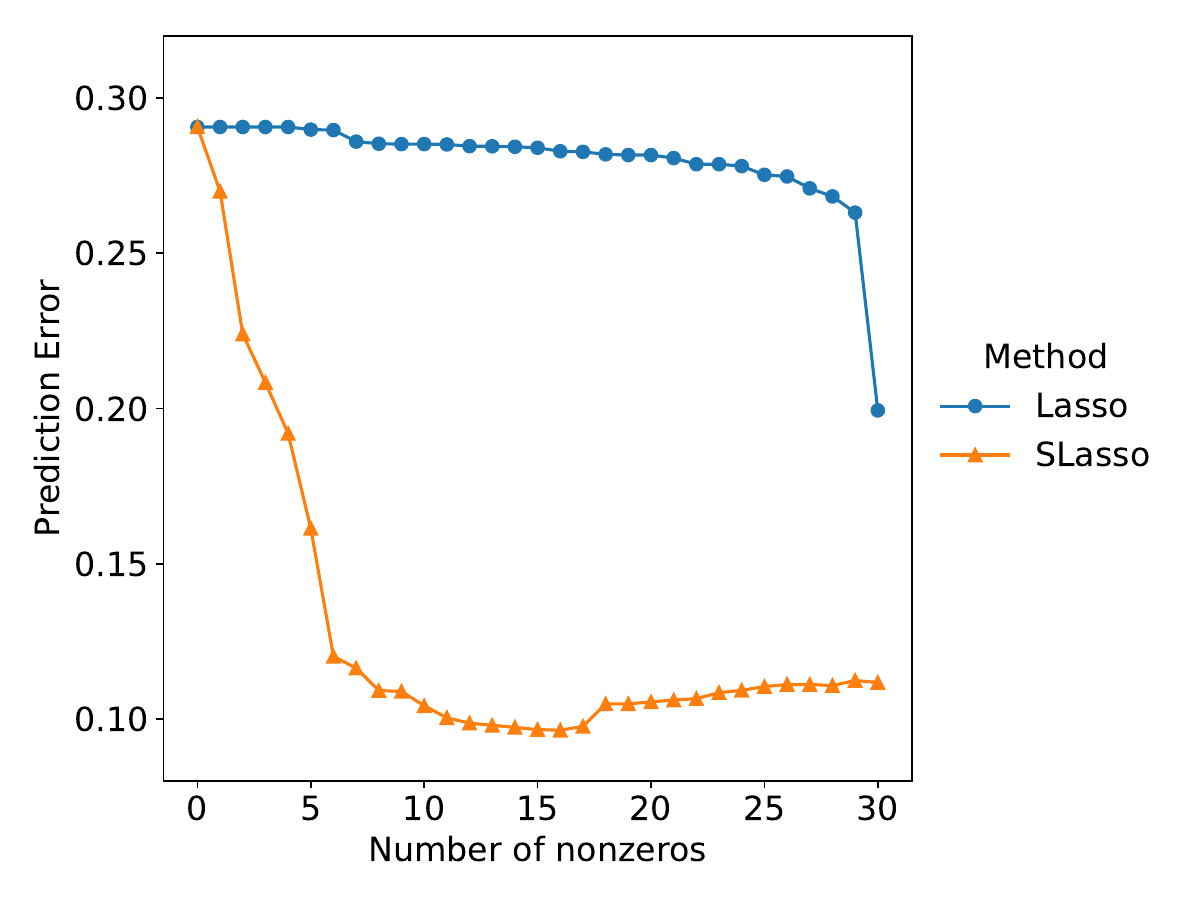}
    \caption{Out of sample prediction errors for Lasso and SLasso in a problem setup with $n=100$, $d=30$, and $s=5$, and the results are averaged over 20 replications. The setup uses $f_0(\boldsymbol{x})=\|\boldsymbol{x}\|_2^2$ with equally-spaced support and correlation level 0.3. Number of nonzeros refers to the number of nonzero subgradients in the respective estimated models.}
    \label{fig: error}
\end{figure}

Figure \ref{fig: error} illustrates the performance of SLasso and compares it with the standard Lasso applied to CNLS. For each fixed level of model sparsity (number of nonzero selected subgradients), there may be multiple tuning parameters of Lasso that achieve this sparsity level. We select the tuning parameter with the optimal out of sample prediction error (similarly for the tuning parameter of SLasso). We see that, across all levels of model sparsity, SLasso dominates Lasso uniformly 
in terms of predictive performance.\footnote{Note that the actual sets of selected variables usually differ for SLass and Lasso, even at the same sparsity level.} In addition, we found that selecting a sparse model with the $\ell_1$-norm is more difficult than using the $\ell_1/\ell_{\infty}$-norm, which has also been observed in previous works. For example, \citet{dai2023variable} found that the Lasso estimator in \eqref{lasso} cannot completely screen out the irrelevant variables in most cases. \citet{xu2016faithful} observe similarly that attempting to select variables by regularizing the subgradients with the standard Lasso is ineffective. On the other hand, like traditional Lasso type methods, SLasso may also excessively penalize the CNLS model, leading to small subgradients for relevant variables and underfitting. For example, in Figure \ref{fig: error}, SLasso still obtains a relatively large prediction error of 0.161 even when it selects the correct number (5) of variables. Using a smaller $\lambda$ improves the predictive performance substantially (e.g., to below 0.1), but also increases the number of nonzero subgradients. This behavior of SLasso suggests that the particular regularization level $\lambda$ that achieves optimal model selection performance may not coincide with the level of $\lambda$ that achieves optimal predictive accuracy. To balance variable selection and predictive performance, in the next section, we extend our framework by proposing a relaxed version of SLasso, which introduces separate tuning parameters for variable selection and model shrinkage.

\subsection{A relaxed version of structured Lasso}\label{subsec: relax}
It has long been known that Lasso type methods could lead to sub-optimal model performance in terms of predictive accuracy \citep{fan2001variable}. Indeed, as demonstrated in Figure \ref{fig: error}, the optimal choice of tuning parameters $\lambda_k$ in SLasso may not result in optimal out of sample prediction performance in a high-dimensional setting. Essentially, the $\ell_1/\ell_{\infty}$-norm used in SLasso \eqref{cnls-slasso} has two effects: variable selection and model shrinkage. On one hand, the sparsity-inducing regularization effect enforces a certain set of subgradients to be zero, thus screening out irrelevant variables from the estimated models. In addition, it has shrinkage effects on the \emph{selected} variables, by also penalizing the $\ell_\infty$-norm of the corresponding subgradients. However, the $\ell_1/\ell_{\infty}$-norm may overly penalize the subgradients of the SLasso model and lead to an underfitting problem. A similar problem has been studied for the standard linear regression Lasso in \cite{meinshausen2007relaxed}. 

To address this underfitting problem, we propose a relaxed version of SLasso for CNLS problems by introducing a new tuning parameter $\gamma$, which regulates the amount of regularization for model shrinkage \emph{separately} from model selection. Let $\hat{f}_{slasso}$ denote the SLasso estimator constructed as in \eqref{estimator} using the solution of problem \eqref{cnls-slasso} with parameters $\lambda_k$. Let the associated active set $\mathcal{M}$ be the indices of the selected variables from the SLasso problem \eqref{cnls-slasso}. In other words, for any $k\notin \mathcal{M}$, $\xi_i^k = 0$ for all $i$. Let $\hat{f}_{cnls}$ be the CNLS estimator using the \emph{selected} variables in $\mathcal{M}$ from the SLasso solution. The relaxed SLasso estimator is defined as

\begin{equation}\label{relax}
    \hat{f}_{relax} := \gamma\hat{f}_{slasso} + (1-\gamma)\hat{f}_{cnls},
\end{equation}
where we emphasize that $\hat{f}_{relax}$ relies on two types of tuning parameters: $\lambda_k\geq0$ (for the SLasso problem) and $\gamma\in (0,1]$. When $\gamma=1$, relaxed SLasso reduces to the SLasso method defined in \eqref{cnls-slasso}, while when $\gamma=0$, relaxed SLasso reduces to the CNLS problem \eqref{cnls} using the selected variables from SLasso based on $\lambda_k$.

To illustrate how the relaxed SLasso can address the problem of over-shrinkage, we present the optimization formulation for computing the relaxed SLasso estimator $\hat{f}_{relax}$. Given tuning parameters $\lambda_k\geq0$ and $\gamma \in (0,1]$, consider the following variant of the SLasso problem
\begin{equation}\label{rslasso}
\begin{aligned}
    \min_{\boldsymbol{\xi}_{\mathcal{M}}; \boldsymbol{\theta}} \quad &  \frac{1}{2} \sum_{i=1}^n(y_i-\theta_i)^2 + \gamma \sum_{k=1}^d \lambda_k\|(\xi_{1,\mathcal{M}}^k, \xi_{2,\mathcal{M}}^k, \ldots, \xi_{n,\mathcal{M}}^k)\|_{\infty}\\
    \mbox{\textit{s.t.}}\quad
    & \theta_i + \boldsymbol{\xi}_{i,\mathcal{M}}^T(\boldsymbol{x}_j-\boldsymbol{x}_i)\leq \theta_j, \quad  i,j=1,\ldots,n, \qquad 
\end{aligned}  
\end{equation}
where, importantly, $\boldsymbol{\xi}_{i,\mathcal{M}}$ denotes subgradients that are identically zero outside the active set $\mathcal{M}$, which is obtained by solving the SLasso problem \eqref{cnls-slasso} using the same $\lambda_k$. Therefore, the relaxed SLasso requires first solving \eqref{cnls-slasso} to select the set of relevant variables $\mathcal{M}$, then solving \eqref{rslasso} which enforces the sparsity by only using variables in $\mathcal{M}$. 

\begin{theorem}\label{theorem: equality}
    For given parameters $\lambda$ and $\gamma$, the relaxed SLasso estimator defined in \eqref{relax} can be obtained using the solutions $\boldsymbol{\xi}, \boldsymbol{\theta}$ to \eqref{rslasso}, following the construction in \eqref{estimator}.
\end{theorem}

\begin{proof} See Appendix A.1.
\end{proof}

 Theorem \ref{theorem: equality} shows that the relaxed SLasso estimator can be obtained by solving the convex optimization problem in \eqref{rslasso}. 
Clearly, using a value of $\gamma$ between 0 and 1, the shrinkage effect on the nonzero subgradients in \eqref{rslasso} is \emph{attenuated} compared to the original SLasso problem \eqref{cnls-slasso}. At the same time, \eqref{rslasso} uses the same set of selected variables $\mathcal{M}$ as those selected by \eqref{cnls-slasso}. As a result, the relaxed SLasso can address the over-shrinkage problem of SLasso while retaining its model selection performance. The formulation \eqref{rslasso} of relaxed SLasso can also be viewed as a generalization of the relaxed Lasso of \cite{meinshausen2007relaxed} for linear regression settings. 

One may be interested in the asymptotic properties of the proposed estimators. Although the consistency and convergence rates of the standard CNLS estimator have been studied before \citep{lim2012consistency,lim2014on}, formal results for regularized CNLS estimators are less common. In particular, there are no previous results for the model selection consistency of Lasso type CNLS estimators. In this work, we defer a more systematic study of the asymptotic properties of the SLasso estimator to future works. However, leveraging the structure of the \emph{relaxed} SLasso as a linear combination of the CNLS and the SLasso estimators, we can show that the relaxed SLasso can achieve the same convergence rate as the standard CNLS under reasonable assumptions. Recall that the relaxed SLasso decouples model selection and shrinkage using two separate parameters, namely $\lambda$ and $\gamma$. The key intuition behind our next result is that, by requiring $\gamma \rightarrow 0$ at the appropriate rate, the relaxed SLasso 
is effectively a CNLS estimator using variables selcted by SLasso, and thus will share an asymptotic behavior similar to the CNLS estimator, \emph{provided} that the relevant variables are correctly selected asymptotically.
\begin{theorem}\label{theorem: fast}
Let $\hat{f}_{relax}^{n}(x)$ be relaxed SLasso estimator defined in \eqref{relax} based on tuning parameters $\lambda^n,\gamma^n$, where the SLasso and CNLS estimators belong to the class of convex functions with bounded subgradients. Suppose that $\gamma^{n}=O_{p}(n^{-1/2})$ and 
$\lambda$ is selected such that $\lim\sup_{n\rightarrow\infty}\mathbb{P}(\mathcal{M}^{n}\neq\mathcal{M}_{0})=0$,
where $\mathcal{M}_{0}, \mathcal{M}^n$ are the supports of  $f_{0}(x)$ and $\hat{f}_{relax}^{n}(x)$, respectively, then we have the following convergence rates:

When $d<4$, as $n\rightarrow\infty$,
\begin{align*}
\frac{1}{n}\sum_{i=1}^{n}(\hat{f}_{relax}^{n}(x_{i})-f_{0}(x_{i}))^{2} & =O_{p}(n^{-4/(4+d)}).
\end{align*}

When $d=4$, as $n\rightarrow\infty$, 
\begin{align*}
\frac{1}{n}\sum_{i=1}^{n}(\hat{f}_{relax}^{n}(x_{i})-f_{0}(x_{i}))^{2} & =O_{p}(n^{-1/2}\log n).
\end{align*}

When $d>4$, as $n\rightarrow\infty$, 
\begin{align*}
\frac{1}{n}\sum_{i=1}^{n}(\hat{f}_{relax}^{n}(x_{i})-f_{0}(x_{i}))^{2} & =O_{p}(n^{-2/d}).
\end{align*}
\end{theorem}

\begin{proof} See Appendix A.2.
\end{proof}

The proof of Theorem \ref{theorem: fast} leverages results of \citet{lim2014on} on the convergence rates of CNLS. Note that our assumption on $\lambda^n$ requires that the penalty parameter of SLasso guarantees its model selection consistency. This is because the primary function of the SLasso in the relaxed SLasso is variable selection, rather than consistent estimation. In a standard Lasso penalized regression, the penalty parameter serves both functions. In the relaxed SLasso, estimation consistency is achieved through a separate parameter $\gamma^{n}$, which is assumed to vanish at a rate of $n^{-1/2}$. As a result, in relaxed SLasso $\lambda^{n}$ is constructed solely to better select the relevant variables. This advantage has been confirmed in our simulation results, where we observe that the relaxed SLasso outperforms the SLasso at selecting the relevant variables, in addition to predictive performance. Of course, the model selection consistency of SLasso and relaxed SLasso remains to be studied systematically.  However, previous works on the relaxed Lasso such as \cite{meinshausen2007relaxed} and our simulation results provide evidence that the cross-validation procedures proposed in this work can select the relevant variables accurately in practice. Moreover, Theorem \ref{theorem: fast} also demonstrates that, when variables are correctly selected, using a small (and vanishing) $\gamma$, the relaxed SLasso can effectively address the underfitting problem that SLasso may encounter.

From a practical perspective, we can obtain the relaxed SLasso estimator by performing Algorithm 1, a two-stage procedure. In the first stage, we solve the initial SLasso problem to perform variable selection, resulting in the active set $\mathcal{M}$, by tuning the parameter $\lambda$ defined in \eqref{aslasso}. In the second stage, given the active set $\mathcal{M}$, we solve the relaxed SLasso problem \eqref{rslasso} with the selected variables, whereby SLasso is relaxed towards CNLS by tuning the parameter $\gamma$. That is, relaxed SLasso mitigates the aggressive shrinkage inherent in SLasso by introducing a trade-off between SLasso and CNLS, controlled by the tuning parameter $\gamma$. 

\floatname{algorithm}{Algorithm}
\begin{algorithm}
\caption{Two-stage relaxed SLasso}
\begin{algorithmic}[1]
    \Require $\{(\boldsymbol{x}_i,y_i)\}_{i=1}^n\in \mathbb{R}^d\times\mathbb{R}$, parameters $\lambda\geq 0$ and $\gamma\in (0,1]$,
    \State Solve the SLasso problem in \eqref{aslasso} with $\lambda$ and return the active index set $\mathcal{M}$;
    \State Solve the relaxed SLasso problem in \eqref{rslasso} using $\mathcal{M}$, $\lambda$, and $\gamma$;
    \Ensure Active set $\mathcal{M}$ and Subgradients $\hat{\boldsymbol{\xi}}_{i,\mathcal{M}}$.
\end{algorithmic}
\end{algorithm}

To choose the tuning parameters $\lambda$ and $\gamma$, we again recommend using cross-validation similar to SLasso, but now with a \emph{joint} grid search. Let $\hat{f}_{relax}$ denote the solution obtained by Algorithm 1 using a particular pair of parameters $(\lambda,\gamma)$. We consider a two-dimensional grid of tuning parameters $\lambda\times \gamma = \{\lambda_1,\ldots,0\}\times \{\gamma_1,\ldots,\gamma_r\}$ with $\lambda_i>\lambda_{i+1}$ and $\gamma_j>\gamma_{j+1}$ for all $i,j$. We set $\lambda_1 = \|\boldsymbol{X}^T\boldsymbol{y}\|_{\ell_1/\ell_{\infty}}$ and $\gamma_j \in (0,1]$. Let 
\begin{equation*}
    \textit{loss}(\lambda, \gamma) = \frac{1}{m} \sum_{i=1}^m(\hat{f}_{relax}(x_i)-y_i)^2
\end{equation*}
denote the loss that evaluates the difference between $y$ and $\hat{f}_{relax}$ on a data set of size $m$. We aim to jointly choose the optimal $\lambda$ and $\gamma$ by minimizing out-of-sample $\textit{loss}(\cdot)$. We employ $k$-fold cross-validation, whereby we compute the average loss $\textit{loss}^{avg}(\cdot)$ by randomly splitting the data set into $k$ equal-sized subsets and computing the out-of-sample loss on each fold, using the rest of the folds to compute the estimator. We choose the optimal parameters $(\lambda^*, \gamma^*) = \argmin_{\lambda\times \gamma} \textit{loss}^{avg}(\lambda, \gamma)$. In our simulation study, we observed that five-fold cross-validation works well.

We conclude this section by emphasizing the motivation behind our relaxed SLasso method: using $\ell_1/\ell_{\infty}$-norm in SLasso may impose an unnecessarily large amount of shrinkage on the selected variables and leads to underfitting. As a result, it may be difficult to choose a suitable tuning parameter $\lambda$ that enforces a sparse model with low prediction errors. On the other hand, the relaxed SLasso isolates the two effects, namely, variable selection and model shrinkage, by utilizing a two-stage procedure that controls these two effects through two separate tuning parameters. In this way, it can select a sparse model while mitigating the risk of underfitting. Simulation results in this work provide evidence that the two tuning parameters can indeed balance those two effects and achieve better performance than SLasso (see the next section and Appendix B.1).

\section{Monte Carlo simulations}\label{sec: mc}

\subsection{Setup}\label{subsec: setup}
Given $n$ (number of observations), $d$ (problem dimensions), $s$ (sparsity level), $\rho$ (correlation level), and signal-to-noise ratio (SNR), we illustrate our methods using  simulations based on  the following model: 
\[y_i = f_0(\boldsymbol{x}_{i}) + \varepsilon_i, \quad i=1,\ldots,n,\]
where $f_0$ is the true regression function. We generate $\boldsymbol{x}_i=(x_i^1,\ldots,x_i^d)\in \mathbb{R}^d$ from a multivariate normal distribution with zero mean and correlation matrix $\Sigma$, where $\Sigma\in \mathbb{R}^{d\times d}$ has entry $(i, j)$ equal to $\rho^{|i-j|}$, $1\leq i,j\leq d$, for some correlation $\rho$.\footnote{Note that when $\rho=0$, the variables are independent and identically distributed (i.i.d.), and the higher $\rho$ is, the larger the correlation among the variables.} We draw $\varepsilon_i$ from the normal distribution $N(0, \sigma^2)$, where $\sigma$ is determined by  $\text{SNR}=\frac{\text{Var}(f_0)}{\sigma^2}$. We note that the higher the value of SNR is, the lower the data noise level. We consider the following two types of data generation processes (DGPs).

Type I: We use a quadratic form as the true regression function (see, e.g., \citealp{bertsimas2021sparse}),
\[f_0(x_{i}^k)=\sum_{k\in S}(x_i^k)^2\]
where the true support set $S$ is selected from $\{1,\ldots,d\}$ with equal space and support set size is $|S|=s$. In Type I, all the nonzero variables have the same magnitude.

Type II: We consider a weighted combination of $s=5$ relevant variables $x_i^k$ for $k\leq 5$,
\[f_0(x_i^k)=\sum_{k=1}^5 \beta_k(x_i^k)^2\]
where the weights $\beta_k$, $k=1,\ldots,5$, are set to be $\{5,4,3,2,1\}$, respectively. For Type II, we take into account the relative importance of variables using distinct weights.

\textit{Selecting the tuning parameters}. To choose the optimal tuning parameters, tuning is performed by minimizing prediction error with five-fold cross-validation on a separate validation set of size $N$; see, e.g., \cite{hastie2009elements}. Lasso and SLasso are tuned over 50 values of $\lambda$ ranging from $\lambda_{max}=\|\boldsymbol{X}^T\boldsymbol{y}\|_{\ell_1/\ell_{\infty}}$ to a small fraction of $\lambda_{max}$. For relaxed SLasso, 10 equally-spaced values of $\gamma\in[0,1]$ are used in a grid search, resulting in 500 candidate tuning parameter pairs $(\lambda, \gamma)$ in total. 

In the following experiments, we use the standard solver MOSEK (version 9.2.44) within the Python/CVXPY package to implement optimization problems. All computations are performed on Aalto University’s high-performance computing cluster Triton with Xeon @2.8 GHz, 10 CPUs, and 8 GB RAM per CPU. The source code and data are available at the GitHub repository (\url{https://github.com/zhiqiangliao/HDCR}).

\subsection{Evaluation metrics}
Let $\hat{f}(\cdot)$ denote the estimated convex function constructed using \eqref{estimator} based on training data of size $n$. We draw $N=1000$ independent test data $\{\boldsymbol{x}_i, y_i\}_{i=1}^N$ (see Section \ref{subsec: setup} for detailed setup) with the same support set as the training set. We then consider the following evaluation metrics:

\textit{Prediction error}: the prediction error (or relative accuracy) metric is defined as \citep{bertsimas2016best}
\begin{equation*}
    \text{Prediction error} = \sum_{i=1}^N |\hat{f}(\boldsymbol{x}_i)-f_0(\boldsymbol{x}_i)|^2 /\sum_{i=1}^N |f_0(\boldsymbol{x}_i)|^2,
\end{equation*}
where $\hat{f}(\cdot)$ denotes the estimated convex function and $f_0(\cdot)$ is the true convex function (e.g., $f_0 = \sum_{k\in S^*} (x_i^k)^{2}$ for the Type I DGP). A perfect score is 0 (e.g. when $\hat{f}(\cdot)\equiv f_0(\cdot)$), while a larger prediction error indicates worse predictive performance.\footnote{We note that \citet{dai2023variable} measured the ``in-sample'' prediction errors, whereas our prediction error metric is ``out-of-sample'', where we compute the average error over a separate independent test data set of size  $N=1000$ instead of the training samples $\boldsymbol{x}_i$, $i=1,\ldots,n$. }

\textit{\# nonzeros}: number of nonzero subgradients in the estimated convex regression model $\hat{f}(\cdot)$. 

\textit{F-score}: unlike the simple \# nonzeros, this metric evaluates variable selection accuracy. This accuracy metric is also studied in \citet{hastie2020best}, and is defined as
\begin{equation*}
    \text{F-score} = \frac{2}{\text{recall}^{-1}+\text{precision}^{-1}},
\end{equation*}
where recall and precision are defined as follows.  
Let $S\subseteq\{1,\ldots,d\}$ be the true support of $f_0$ and $\hat{S}$ the estimated support set, i.e., the support of $\hat f$. The precision and recall given $S$ can be defined as
\begin{equation*}
    \text{recall}=\frac{|S\cap \hat{S}|}{|\hat{S}|} \quad \text{and} \quad 
    \text{precision}=\frac{|S\cap \hat{S}|}{|S|}
\end{equation*}
respectively. A perfect value of the F-score is 1.

\textit{$R^2$}: proportion of variance explained by the estimator $\hat{f}(\cdot)$, that is,
\begin{equation*}
    R^2 = 1-\frac{\sum_{i=1}^N(y_i-\hat{f}(\boldsymbol{x}_i))^2}{\sum_{i=1}^N(y_i-\bar{y})^2}
\end{equation*}
where $\bar{y}=\frac{1}{N}\sum_{i=1}^N y_i$, which is the mean of the observed data. A perfect score is one and the null score is zero.

\textit{Jaccard similarity score}: the Jaccard similarity score describes the similarity between the sets of selected variables and, for two sets $S_1$ and $S_2$, is defined as
\begin{equation*}
    \text{Jaccard similarity score} = \frac{S_1\cap S_2}{S_1\cup S_2}.
\end{equation*}
To examine the \emph{stability} of each variable selection method, we generate 50 independent data sets and use the variable selection methods to estimate the support set for each data set. For each variable selection method and for every pair of estimated support sets based on two distinct data sets, we calculate the Jaccard similarity score between them. We then average the Jaccard similarity scores across all pairs of generated data sets for each method; also see the recent work of \cite{donnelly2023rashomon}. A perfect score of 1 means that the method selects the exact same set of variables across all 50 independent data sets, i.e., its variable selection is stable. Smaller average Jaccard similarity indicates more variability of a variable selection method across random data sets.

\subsection{Variable selection performance}
\label{subsec:variable-selection-simulation}
We present comparisons of the finite sample performance of the proposed SLasso and relaxed SLasso methods with four other variable selection techniques popular in the literature. Two methods with shape constraints used in the comparisons are the Lasso CNLS method proposed by \cite{lee2020lasso} and the elastic net (EN) method used by \cite{duras2023using}, while two methods without shape constraints are the linear Lasso (LL) proposed by \cite{tibshirani1996regression} and the modern machine learning algorithm, extreme gradient boosting (XGBoost), developed by \cite{chen2016xgboost}.

\begin{table}[H]
    \centering
    \caption{Average \# nonzeros and F-score of six variable selection methods for data types I and II as $n$ and $d$ vary. The setup fixes SNR to 3, variable correlation level to 0.3, and true support size $s=5$. The simulation is repeated 50 times.}
    
    \begin{tabular}{l c c c c c c}
    \hline
    \multicolumn{7}{c}{\# nonzeros}\\
    \hline
    & SLasso & Relaxed SLasso & Lasso & EN & LL & XGBoost \\
    \hline
    Type I ($n=100$, $d=10$) & 5.9 & \bftab 5.1 & 7.0 & 6.9 & 9.5 & 5.3 \\
    Type I ($n=500$, $d=50$) & 5.0 & \bftab 5.0 & 40.1 & 36.9 & 28.8 & 5.2 \\
    Type II ($n=100$, $d=10$) & 6.8 & \bftab 4.9 & 6.8 & 6.2 & 8.4 & 4.0 \\
    Type II ($n=500$, $d=50$) & 6.0 & \bftab 4.7 & 38.2 & 31.8 & 31.8 & 3.5 \\
    \hline
    \multicolumn{7}{c}{F-score}\\
    \hline
    & SLasso & Relaxed SLasso & Lasso & EN & LL & XGBoost \\
    \hline
    Type I ($n=100$, $d=10$) & 0.923 & \bftab 0.981 & 0.685 & 0.749 & 0.689 & 0.862 \\
    Type I ($n=500$, $d=50$) & \bftab 1.0 & 0.996 & 0.189 & 0.224 & 0.225 & 0.960 \\
    Type II ($n=100$, $d=10$) & 0.828 & \bftab 0.887 & 0.649 & 0.663 & 0.642 & 0.772 \\
    Type II ($n=500$, $d=50$) & 0.882 & \bftab 0.920 & 0.207 & 0.239 & 0.221 & 0.797 \\
    \hline
    \end{tabular}
    \label{tab: select1}
\end{table}

To evaluate the performance of these methods at variable selection, we compare the evaluation metrics \# nonzeros and F-score on synthetic data sets based on DGP Type I and DGP Type II. We show results for $n\in\{100,500\}$ and $d\in\{10,50\}$ while fixing $s=5$, $\rho=0.3$, and $\text{SNR}=3$. The results are presented in Table \ref{tab: select1}. Evaluation metrics are computed on the $n$ design samples (i.e., training data), and the results are averaged over 50 replications. We observe that our proposed methods, SLasso and relaxed SLasso, achieve the best performance in terms of variable selection. Specifically, relaxed SLasso achieves the number of nonzeros closest to the true support set size $s=5$ and obtains the highest F-score. Traditional variable selection methods for convex regression, such as Lasso and EN, perform poorly at eliminating all irrelevant variables, as does the linear Lasso (LL). XGBoost 
comes closest to our structure Lasso methods, but often fails to select all the relevant variables, especially when the relative importance of relevant variables is heterogeneous (as in DGP Type II). These results highlight the significant improvements in variable selection for convex regression when we explicitly incorporate subgradient structures into the estimation procedure. 

Since our main purpose in developing SLasso is to find a parsimonious model while avoiding overfitting, we examine the out-of-sample performance of these methods in the next section.

\subsection{Out-of-sample performance}
\begin{figure}[H]
    \centering
\includegraphics[width=0.7\linewidth]{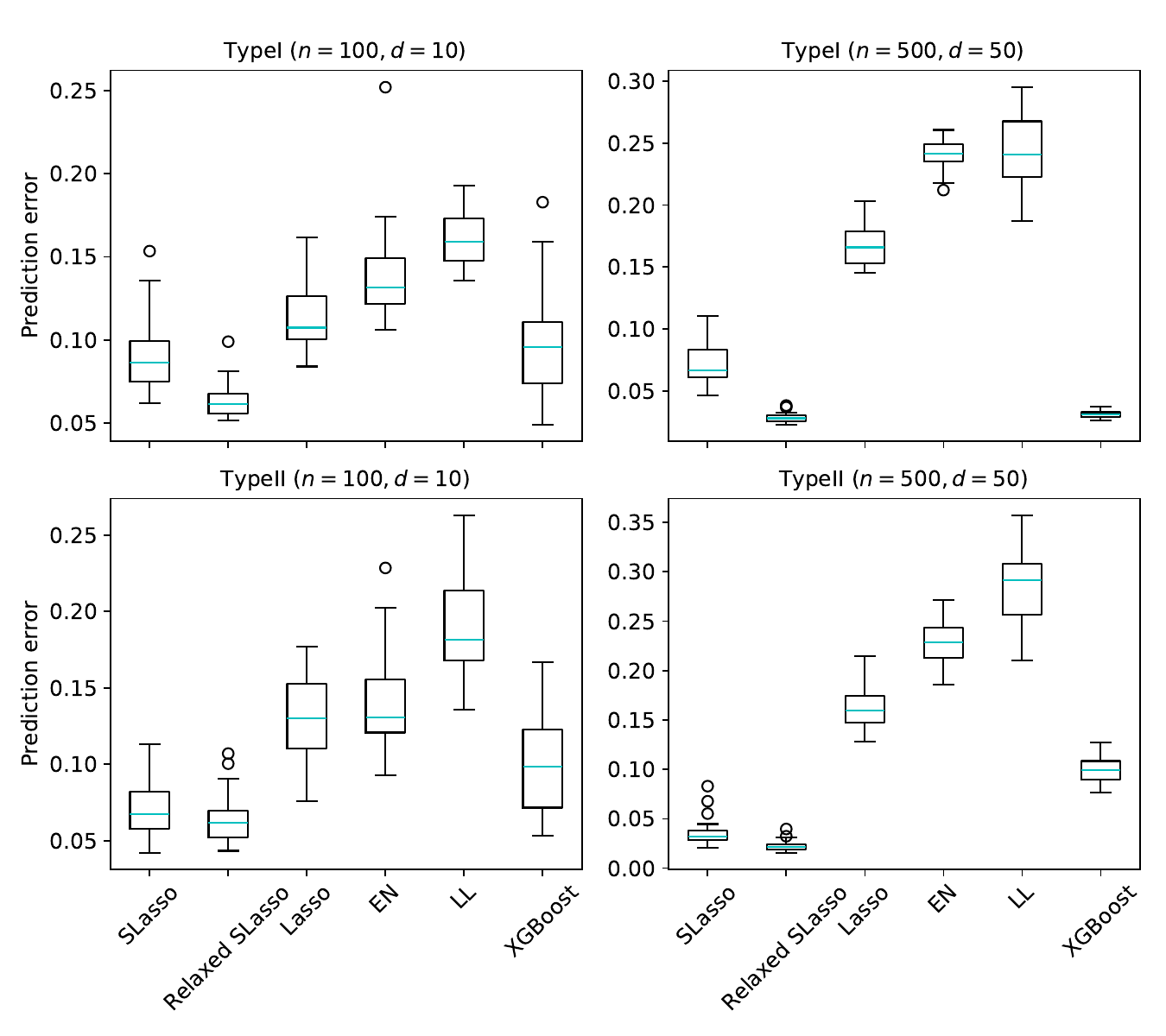}
    \caption{Prediction errors of the six methods in the simulation study with correlated variables where $\rho=0.3$ and $\text{SNR}=3$ for DGPs Type I and Type II as $n$ and $d$ vary.}
    \label{fig: prediction error}
\end{figure}
We use $N=1000$ independently generated observations to test the out-of-sample performance of the sparse models chosen by the six methods. We note that LL and XGBoost are only used for variable selection, and then the CNLS defined in problem \eqref{cnls} is used to fit a convex function with the selected variables. The correlation level is fixed at $\rho=0.3$ and the data noise level is set to $\text{SNR}=3$. Figure \ref{fig: prediction error} summarizes the results using box plots, where the boxes are drawn from the lower to the upper quartile with a line representing the median of the data. The results show the comparative out-of-sample prediction errors of six methods in 50 independent simulations. The two new structured Lasso methods, SLasso and relaxed SLasso, display the best predictive accuracy overall. As expected, relaxed SLasso achieves better out-of-sample performance than SLasso due to its ability to alleviate underfitting. Traditional Lasso and EN obtained relatively small prediction errors in small problem settings ($n=100$ and $d=10$), but their performance deteriorates significantly in larger settings ($n=500$ and $d=50$). As observed earlier, LL and XGBoost may select a model with more variables (i.e., LL) or fewer variables (i.e., XGBoost) than necessary, and thus their out-of-sample performance may suffer from falsely selected variables and the overfitting problem. 

\begin{figure}[H]
    \centering
    \includegraphics[width=0.7\linewidth]{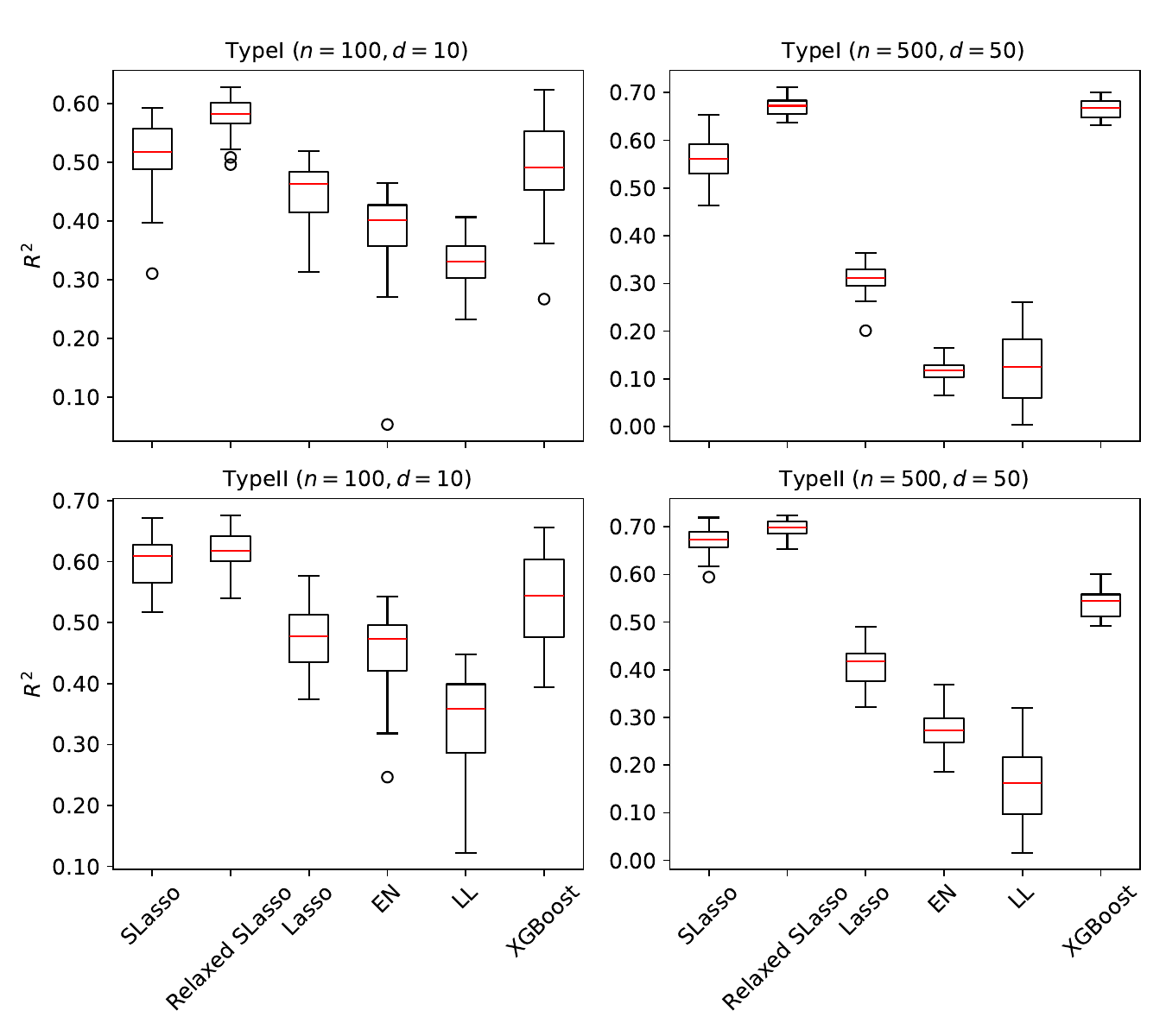}
    \caption{$R^2$ of the six methods in the simulation study with correlated variables where $\rho=0.3$ and $\text{SNR}=3$ for DGPs Type I and Type II as $n$ and $d$ vary.}
    \label{fig: r2}
\end{figure}

Figure \ref{fig: r2} displays the box plots for $R^2$. In general, SLasso and relaxed SLasso are better than Lasso, EN, LL, and XGBoost in terms of explained variance, but SLasso lags behind XGBoost for scenario Type I ($n=500$, $d=50$). We also observe that relaxed SLasso obtains consistent superior performance, achieving the largest $R^2$ value among the six methods. The machine learning method, XGBoost, performs well in large settings ($n=500$, $d=50$) but is still outperformed by relaxed SLasso. In addition, XGBoost performs worse in Type I than in Type II DGP, while relaxed SLasso is consistently competitive in all cases. In a nutshell, relaxed SLasso maintains its good performance under changing DGPs, even compared to modern machine learning methods.

\subsection{Stability of methods}
To evaluate the stability of the proposed methods, we use DGP Type I with $n=100$ and $d=10$. We generate 50 independent data sets with varying data noise level SNR and correlation level $\rho$, where SNR ranges from 0.1 to 7 and $\rho$ varies between 0 and 0.7. 

\begin{figure}[H]
    \centering
    \includegraphics[width=0.8\linewidth]{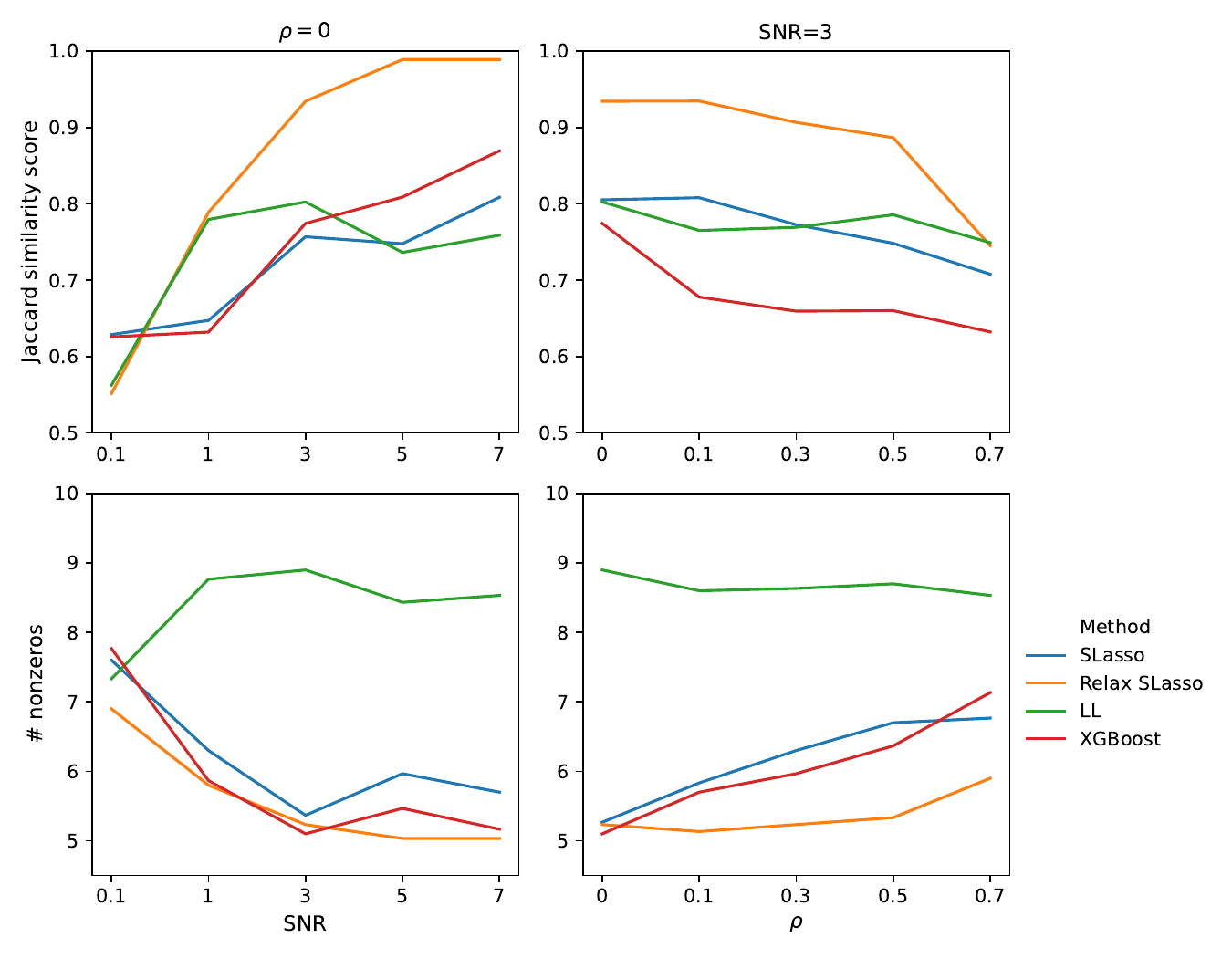}
    \caption{The averaged Jaccard similarity score and \# nonzeros as the data noise level $\text{SNR}$ and correlation level $\rho$ vary for DGP Type I $n=100$, $d=10$, and $s=5$. $\rho$ is fixed at $\rho=0$ for the left two panels; SNR is fixed at $\text{SNR}=3$ for the right two panels. }
    \label{fig: jac}
\end{figure}

We report the Jaccard similarity score and \# nonzeros of four methods. Linear Lasso and EN are excluded because they are inefficient for variable selection (see, e.g., \citealp{xu2016faithful}). Figure \ref{fig: jac} shows the results of the Jaccard similarity scores and \# nonzeros averaged over 50 replications. We observe that relaxed SLasso is generally more stable for variable selection in different runs than other methods. SLasso selects variables more stably than other methods when $\text{SNR}=0.1$, but is outperformed by relaxed SLasso as the SNR increases. On the other hand, when the multicollinearity of the variables is strong (as $\rho$ increases), Figure \ref{fig: jac} shows that relaxed SLasso still dominates SLasso, LL, and XGBoost, while SLasso lags behind LL or XGBoost in terms of stability. In summary, relaxed SLasso also achieves the best performance in terms of variable selection stability when data perturbation and multicollinearity exist. We also observe that relaxed SLasso selects a sparser model than other methods in almost all scenarios, while exhibiting the best stability.

\begin{figure}[H]
    \centering
    \includegraphics[width=0.8\linewidth]{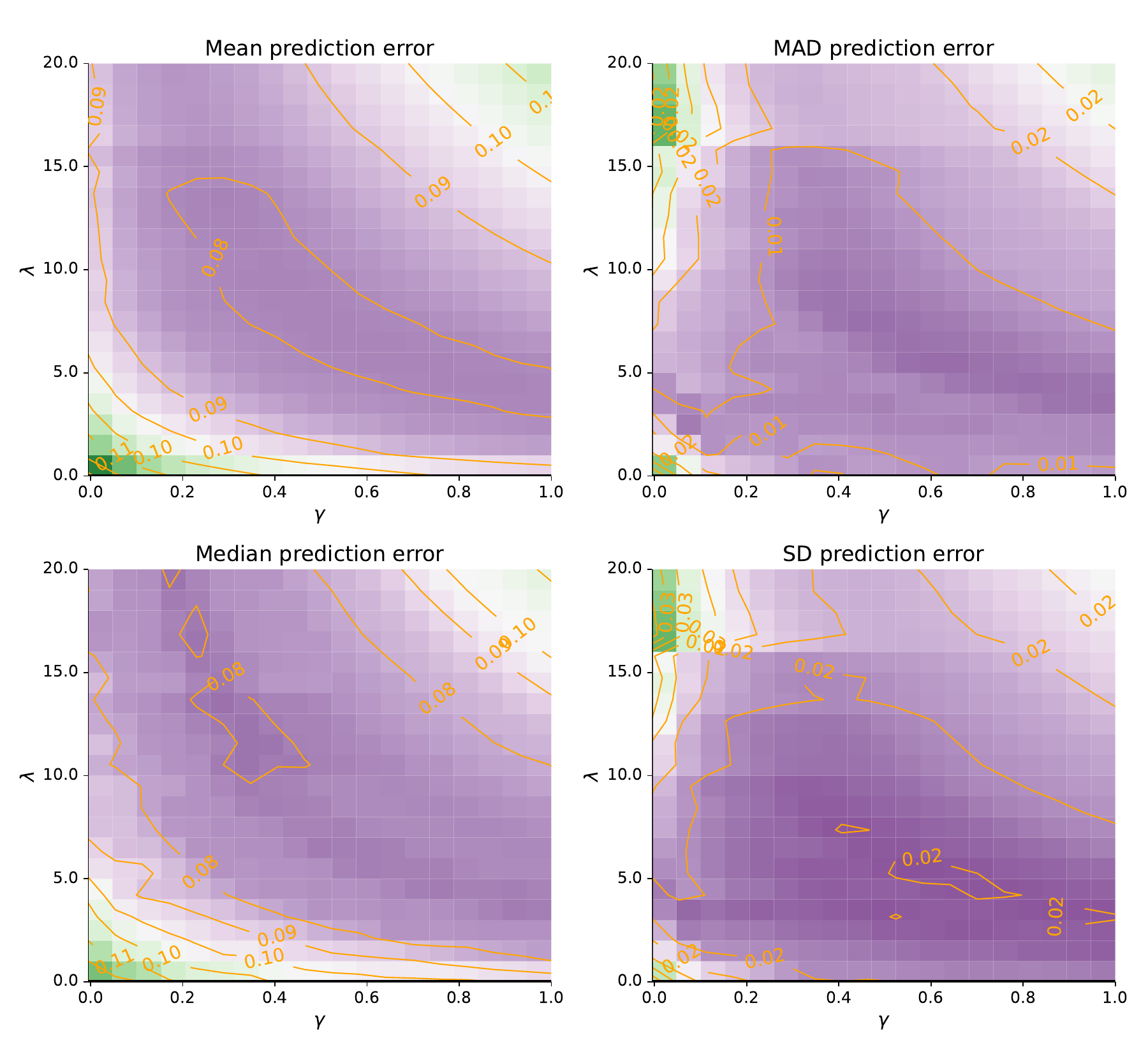}
    \caption{Statistics of prediction errors of the relaxed SLasso with varying $\lambda$ and $\gamma$. 
    }
    \label{fig: stable}
\end{figure}

To demonstrate how the choices of parameters $\lambda$ and $\gamma$ affect the performance of relaxed SLasso, we use the same simulation scheme as before, fixing the SNR at 3 and correlation $\rho$ at 0.3. We compute the out-of-sample performance of the models selected by relaxed SLasso and report the prediction errors. Figure \ref{fig: stable} illustrates the mean, median, median absolute deviation (MAD), and standard deviation (SD) of out-of-sample predictive errors. We observe that the different measures of out-of-sample errors are very stable for almost all ($\lambda$, $\gamma$) pairs. The mean out-of-sample prediction errors with large $\lambda$ and $\gamma$ remain small and are robust to the choices of these two parameters. This robustness property of relaxed SLasso facilitates the tuning of the parameters $\lambda$ and $\gamma$ in practice. Overall, the numerical results in this section demonstrate that our proposed variable selection method, relaxed SLasso, maintains stability across reasonable data perturbations, and the selected models can deliver reliable predictive performance and are robust to tuning parameter choices.

\subsection{Computational performance}
\begin{figure}
    \centering
    \includegraphics[width=0.6\linewidth]{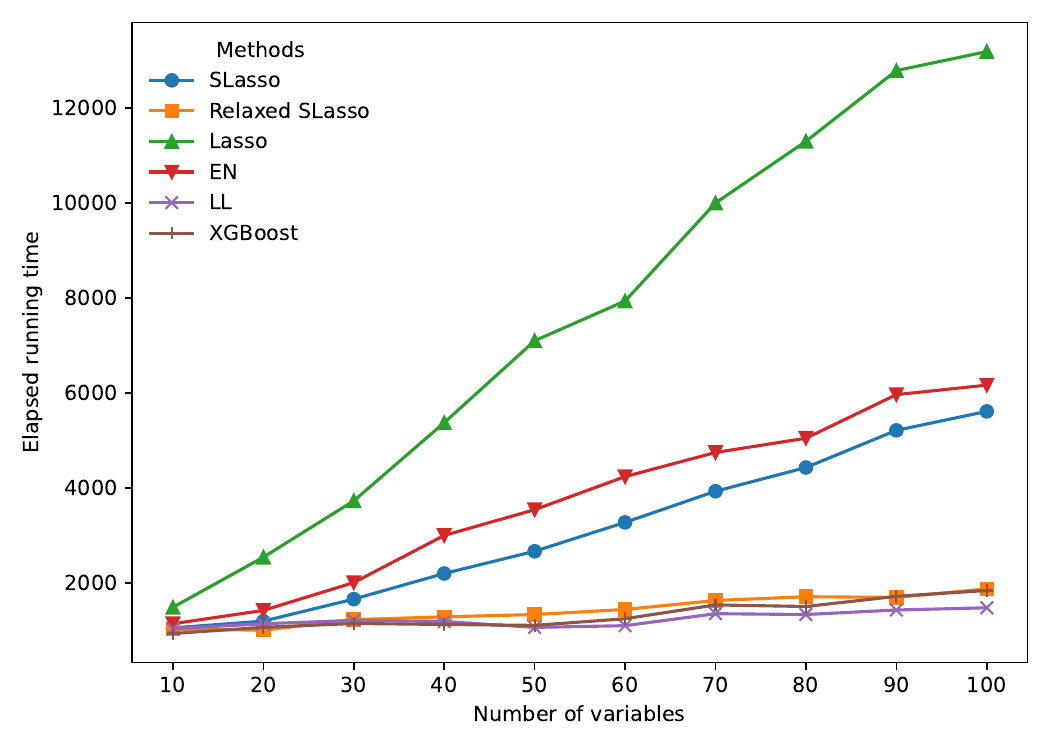}
        \caption{Running time (in seconds) of the six methods given $n=500$ and as $d$ varies from 10 to 100. All methods search over a total of 20 tuning parameter values and select the best sparse model for the comparison.}
    \label{fig: time}
\end{figure}
Next, we study the computational performance of our proposed methods by comparing them against the methods described in Section \ref{subsec:variable-selection-simulation}. The data is generated i.i.d. from DGP Type I with $n=500$, $\text{SNR}=3$ and $\rho=0.3$. To compare the computational performance of the six methods, we vary the number of variables $d$ from 10 to 100 for these methods. We perform cross-validation for each method over 20 values of the tuning parameters for this computational benchmark exercise.

As shown in Figure \ref{fig: time}, relaxed SLasso, LL, and XGBoost scale almost linearly and are much faster than the other three methods. On the other hand, the traditional Lasso is surprisingly slow. Given that the standard Lasso is inefficient for variable selection, the models it selects tend to include far more variables than necessary when the true model is sparse, which in turn leads to poor computational performance. In contrast, the proposed SLasso and relaxed SLasso are able to select much sparser models that overlap or coincide with the true support of the model, and can reduce the overall computation time significantly compared to Lasso. Furthermore, by using a two‑stage procedure, relaxed SLasso reduces the computational burden in the second stage because it uses a sparse selected model when running CNLS. Although LL and XGBoost outperform slightly in some cases, relaxed SLasso remains competitive in terms of computation time compared to both the linear and machine learning methods. However, as is the case for solving the standard CNLS problem, commercial solvers used in our experiments also may not scale well and become computationally expensive when $n>500$ due to the large number of shape constraints ($n^2$). Recent research has examined the computational aspects of CNLS problems, and may be an independent topic of interest for our proposed structured Lasso CNLS methods. See, e.g., \citealp{mazumder2019a}, \citealp{bertsimas2021sparse}, \citealp{lin2022augmented}, and \citealp{chen2024subgradient}.

\section{Application}\label{sec: app}
In this section, we empirically evaluate the performance of the proposed methods in terms of variable selection and predictive accuracy, using data provided by SEMI from 2016 to 2019. SEMI annually collects technical, accounting, and market data from Swedish electricity distribution system operators (DSOs). Based on these data, SEMI aims to estimate a cost function that can be used in efficiency analysis and energy regulation. However, as discussed in SEMI \citeyear{swedish2021}, there is severe multicollinearity between production variables. In this paper, we use the proposed variable selection techniques to tackle this problem and compare their performance with other methods in the literature. To evaluate the estimated cost functions, we consider the test error and \# nonzeros metrics in this empirical application.

\subsection{The regulation of Swedish electricity distribution system operators}
The DSOs usually enjoy a natural local monopoly due to expensive construction fees. In many countries, energy regulators aim to monitor electricity distribution networks, reduce their local monopoly power, and encourage them to reduce the total operational cost \citep{kuosmanen2012stochastic}. The regulation of electricity distribution networks typically involves two procedures: cost function estimation and cost efficiency analysis. Estimating an appropriate cost function is the fundamental and crucial step in the energy regulation problem. In this paper, we apply the proposed methods to fit a cost function, which is assumed to be convex and non-decreasing, using the data provided by SEMI.\footnote{The SLasso-CNLS estimator with non-decreasing constraints is defined in Appendix B.2.} In the second-stage analysis, SEMI could use the efficiency analysis method to calculate the level of economic efficiency for each DSO (see, e.g., \citealp{kuosmanen2012stochastic}). This paper focuses on the task in the first stage: estimation of cost functions. 

After excluding some small DSOs that are outliers, we used 144 out of 154 Swedish distribution firms \citep{duras2023using}. The number of input and output variables is 25, and they have been commonly used in the literature (see, e.g., \citealp{kuosmanen2020capital,duras2023using}). Like the SEMI's current model, we use the total cost (\textit{TOTEX}) as the single output variable. The input variables ($\boldsymbol{x}$) include potentially relevant production variables currently used in the SEMI model, as well as network and contextual variables not considered by SEMI. They are described in detail in Appendix B.3. In addition, we also added three green energy production variables: \textit{Wind Power}, \textit{Local Energy}, and \textit{Solar Power}. By considering these green energy variables, we can investigate the impact of renewable energy sources on the total operational cost of DSOs. We thank Dr. Kristofer Månsson for providing the renewable energy data. Inspecting the data, we notice that some variables dominate others in magnitude, indicating that these variables are likely to dominate the function estimation. To eliminate the influence of this scale heterogeneity, we standardize the input and output variables to achieve a more reliable estimation of the cost functions.

\subsection{Variable selection and estimates}
To illustrate the performance of our methods, we estimate a sparse cost function model using the proposed variable selection methods based on the data collected from Swedish DSOs. We compare our methods with SEMI's current model, which uses seven manually selected variables. SEMI discussed with several Swedish DSOs to determine which variables are important for estimating the operational cost. In contrast, we aim to select the relevant variables through a data-driven procedure. We note that the model of \citet{duras2023using}, which is also considered for comparison, applies the traditional Lasso for variable selection. 

To select the optimal tuning parameters $\lambda$, we use the five-fold cross-validation described in Section \ref{sec: mc}. In particular, we consider a series of tuning parameters, where $\lambda$ takes equally spaced 50 values over $[1,\ldots,100]$. Since the relaxed SLasso is also considered, an additional parameter $\gamma$ is tuned over 10 equally-spaced values between 0 and 1.

The results are presented in Table \ref{tab: estimates}.\footnote{Blank space denotes that the variable has been eliminated, and checkmarks indicate the selected variables.} ``SEMI'' represents the current model used by SEMI, SLasso and relaxed SLasso are the proposed methods, and ``Lasso'' and ``EN'' are the variable selection methods considered by \cite{duras2023using} for convex regression problems. ``XGBoost'' is also included as a benchmark, representing modern machine learning methods, and ``LL'' denotes the standard Lasso regression method. We observe that SLasso selects a sparse model with eight variables, while relaxed SLasso makes the model even sparser by eliminating the variable \textit{HV Subs}. Importantly, while relaxed SLasso selects the same number (7) of variables as SEMI's model, the latter depends solely on production variables. In contrast, relaxed SLasso only selects 4 of the 7 production variables used by SEMI, while adding three variables relevant to electricity networks (\textit{ILVOL}, \textit{LVUL}, and \textit{HVUL}). This change results in significant improvements in predictive power (see Section \ref{subsec:predictive-performance}). Duras' method results in a model with more variables than SEMI's model and includes more variables than are typically considered by Swedish DSOs in practice (see the report SEMI \citeyear{swedish2021}). In contrast, SLasso and relaxed SLasso retain most variables related to low-voltage, such as \textit{LV Energy} and \textit{LV Subs}, while eliminating high-voltage variables, such as \textit{HV Energy} and \textit{HV Subs}. This aligns with the fact that DSOs primarily deliver low-voltage electricity to households and industrial users. Baseline methods without shape constraints, LL and XGBoost, select a sparse model but include contextual variables that other methods do not.

Our results also provide evidence on whether renewable energy production increases the monetary cost (or shadow prices) of Swedish DSOs. Using five variable selection techniques, we examine the effect of the three renewable energy variables--\textit{wind power}, \textit{local energy}, and \textit{solar power}--on the total cost of DSOs. Table \ref{tab: estimates} shows that none of the methods except LL identify the renewable energy variables as relevant contributors to the increase in total cost. We contend that delivering distributed energy resources (e.g., wind, solar, batteries) to local users may not increase the total operational costs of Swedish DSOs. 

\begin{table}
\small
\centering
\begin{threeparttable}
\caption{List of variables selected by using different methods}
    \begin{tabular}{l c c c c c c c}
    \hline
   Variables & SEMI & SLasso & Relaxed SLasso &  Lasso  & EN & LL & XGBoost \\
    \hline
    \multicolumn{8}{c}{Production variables}\\
    LV Energy     & \checkmark & \checkmark & \checkmark & \checkmark & \checkmark & \checkmark & \checkmark \\
    HV Energy     & \checkmark &            &            &            & \checkmark & \checkmark & \checkmark \\
    LV Subs       & \checkmark & \checkmark & \checkmark & \checkmark & \checkmark & \checkmark & \checkmark \\
    HV Subs       & \checkmark & \checkmark &            & \checkmark & \checkmark && \checkmark \\
    T-Power       & \checkmark & \checkmark & \checkmark & \checkmark & \checkmark & \checkmark & \checkmark \\
    S-Power       & \checkmark & \checkmark & \checkmark & \checkmark & \checkmark & \checkmark & \checkmark \\
    Networks      & \checkmark &            &            & \checkmark & \checkmark & \checkmark & \checkmark \\
    Wind Power    &            &            &            &            &            & \checkmark &             \\
    Local Energy  &            &            &            &            &            &&             \\
    Solar Power   &            &            &            &            &            &\checkmark &             \\ 
    \multicolumn{8}{c}{Network variables}\\
    ULVOL         &            &            &            &            & \checkmark & \checkmark &            \\
    ILVOL         &            & \checkmark & \checkmark & \checkmark & \checkmark & \checkmark & \checkmark \\
    LVUL          &            & \checkmark & \checkmark & \checkmark & \checkmark & \checkmark & \checkmark \\
    UHVOL         &            &            &            &            & \checkmark &&             \\
    IHVOL         &            &            &            &            &            &&             \\
    HVUL          &            & \checkmark & \checkmark & \checkmark & \checkmark & \checkmark & \checkmark  \\ 
    \multicolumn{8}{c}{Contextual variables}\\
    Temp        &              &            &            &            &            &&             \\
    Industry    &              &            &            &            &            &&             \\
    Public      &              &            &            &            &            & \checkmark & \checkmark  \\
    Household   &              &            &            &            &            && \checkmark  \\
    Agriculture &              &            &            &            &            & \checkmark & \checkmark  \\
    Commerce    &              &            &            &            &            & \checkmark & \checkmark \\
    Density     &              &            &            &            &            &&             \\
    Growth      &              &            &            &            &            &&             \\
    \hline
    \end{tabular}
\label{tab: estimates}
\end{threeparttable}
\end{table}

\subsection{Predictive performance}
\label{subsec:predictive-performance}
To assess the predictive performance of different methods, we compare test errors and \# nonzeros of those methods based on the SEMI data set. We randomly split the data set into five folds, train all the models on four-fold data, and, for each model, calculate test errors and \# nonzeros on the remaining one-fold data. 
That is, given 144 observations $\{(\boldsymbol{x}_i,y_i)\}_{i=1}^{144}$, we train model $\hat{f}(\cdot)$ on 114 data points and compute the test error on the remaining 30 data points:
\begin{equation*}
    \text{Test error} = \frac{1}{30}\sum_{i=1}^{30} |\hat{f}(\boldsymbol{x}_i)-y_i|^2.
\end{equation*}
A large test error indicates that the estimator has poor predictive performance. The results are averaged over ten random splits. Recall that SEMI uses seven manually selected variables for their current model, using a procedure based on interviews conducted with several Swedish DSOs and statistical tests by researchers (see the report issued by SEMI \citeyear{swedish2021}). Therefore, we obtain the prediction results for SEMI's model using the CNLS estimator in \eqref{cnls} with the seven variables. We perform a similar procedure for XGBoost and LL because they are only used for variable selection in this application. For other methods, including SLasso, relaxed SLasso, Lasso, and EN, we are allowed to select the variables and estimate the model coefficients simultaneously.

\begin{figure}[H]
\centering
    \includegraphics[width=0.7\linewidth]{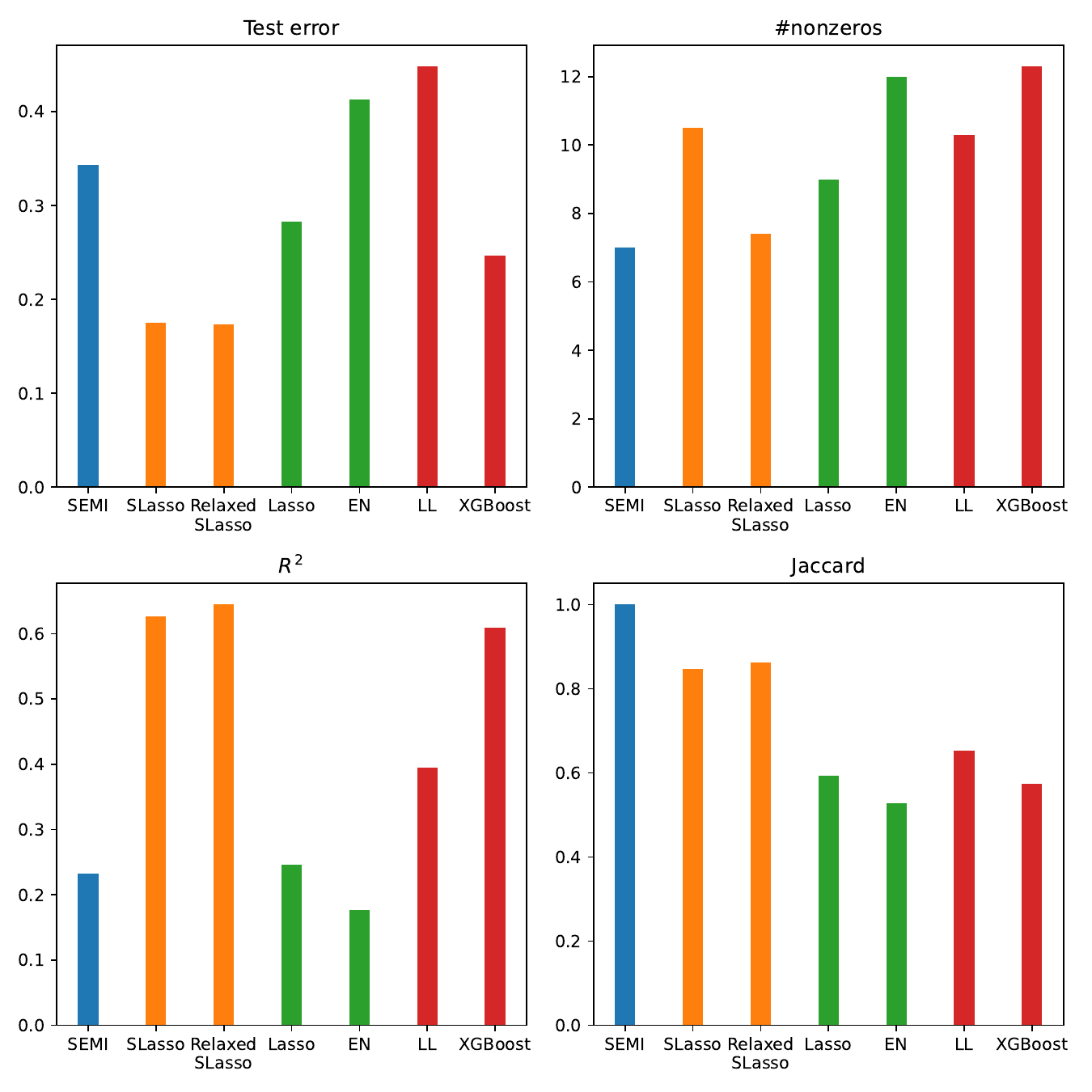}
    \caption{Test errors, the number of nonzeros (\# nonzeros), $R^2$, and Jaccard similarity scores (Jaccard) using various variable selection methods based on the SEMI data set.}
    \label{fig: semi_models}
\end{figure}

Figure \ref{fig: semi_models} shows that SEMI results in large average test errors, indicating poor predictive performance on unseen data. We note that there is multicollinearity among the production variables (see the discussion in SEMI \citeyear{swedish2021}). As a result, models with too many production variables may be unreliable and thus yield high prediction errors. Duras' EN model performs poorly in terms of both prediction and variable selection, yielding a test error of around 0.4 and approximately 12 nonzero coefficients. We observe that SLasso and relaxed SLasso achieve better performance and yield similar test errors, with test errors near 0.17. As shown in Figure \ref{fig: semi_models}, relaxed SLasso produces an even sparser model than SLasso, resulting in approximately 7 nonzero variables, which is close to the number determined by DSOs in the report (SEMI, \citeyear{swedish2021}). We expect that Swedish DSOs will find the model selected by the relaxed SLasso easier to interpret than models produced by other methods. XGBoost performs fairly well in terms of test error, but loses its advantage when it comes to variable selection: when implementing XGBoost on the SEMI data set, we find that its performance is unstable for variable selection (see results on Jaccard similarity in Figure \ref{fig: semi_models}). Moreover, the performance of this machine learning method depends on the availability of a relatively large amount of data.

Given the ten random splits generated beforehand, we selected the variables using different methods for each training set. We then computed the Jaccard similarity scores between the selected support sets and averaged the results across the ten splits. Figure \ref{fig: semi_models} illustrates the Jaccard similarity scores for six methods. Note that the variables in SEMI are predetermined by DSOs, which leads to a Jaccard similarity score of 1. For the other variable selection techniques, we observe that SLasso and relaxed SLasso achieve the highest stability, with scores around 0.8. Our experiments confirm the results in \cite{duras2023using} that Lasso performs better than EN in a high-correlation setting. However, they are both outperformed by the proposed methods in scenarios where a sparse model is desired. We conclude that relaxed SLasso obtains the smallest test error and chooses the most parsimonious models with high stability, suggesting that it successfully identifies the important variables that influence the total operational cost of DSOs.

\section{Conclusions}\label{sec: concl}
In this paper, we develop a framework for variable selection in convex nonparametric least squares (CNLS) problems. Given the distinct structure of subgradients in CNLS problems, the traditional Lasso method that penalizes all entries of the subgradient matrix is inefficient at estimating a sparse model, as it shrinks estimated subgradients but cannot completely zero out all entries corresponding to the irrelevant variables. We introduce structured Lasso methods, namely SLasso and relaxed SLasso, for variable selection in CNLS regression. The key idea underlying these methods is to combine the use of $\ell_1$-norm and $\ell_{\infty}$-norm to identify the unique column-wise sparsity structure of the subgradient matrix. Compared to SLasso, relaxed SLasso offers the additional benefit of controlling variable selection and model shrinkage separately, thereby enhancing predictive power while preserving model sparsity.

Evidence from the MC study and empirical application demonstrate a number of advantages of our proposed SLasso and relaxed SLasso methods. First, they often significantly outperform existing methods in the literature in terms of both model sparsity (Table \ref{tab: select1}) and predictive accuracy (Figures \ref{fig: prediction error} and \ref{fig: r2}). For regression problems with shape constraints, standard Lasso methods often fail to eliminate all irrelevant variables and suffer from poor predictive accuracy. In contrast, SLasso and relaxed SLasso can zero out the subgradients of the irrelevant variables efficiently and thus achieve sparsity in the selected model. While both methods enjoy superior predictive accuracy, the relaxed SLasso achieves additional accuracy gains compared to SLasso, and consistently outperforms other variable selection methods across different levels of data noise and variable correlation. Second, our proposed methods demonstrate strong stability in terms of the set of selected variables when applied to random data instances from the same generating distribution (Figure \ref{fig: jac}), making it less sensitive to randomness in the data and thus more reliable in practice. Third, our methods also exhibit robustness across a range of tuning parameter values (Figure \ref{fig: stable}), reducing the need for extensive grid search refinement. Fourth, the proposed methods are among the least computationally intensive methods considered in this paper (Figure \ref{fig: time}), with relaxed SLasso only slightly slower than the standard linear Lasso. Last but not least, the empirical application further validates the appeal of our variable selection methods for CNLS. On the SEMI data, they are able to eliminate several production variables from the default model used by SEMI based on manual selection, in favor of electricity network variables previously not considered (Table \ref{tab: estimates}), achieving more than 50\% improvements in terms of predictive accuracy over SEMI's model. In contrast, other methods from the literature fail to achieve the same level of sparsity, are less stable in the selected set of variables, and have worse predictive accuracy (Figure \ref{fig: semi_models}). Overall, our results suggest that relaxed SLasso is the most reliable method for estimating sparse models for CNLS problems, and produces the most accurate predictive models.

Going forward, we expect that the convexity constraints considered in this paper can be relaxed in future works, for example, by considering quasiconvex or s-shaped functions in high-dimensional settings. A systematic investigation of the asymptotic properties of SLasso and relaxed Lasso, particularly model selection consistency, would also complement the strong empirical performance of the proposed methods. Designing customized optimization routines to efficiently solve shape-constrained regression problems, including the standard CNLS and our proposed methods, remains an important topic with practical value.

\section*{Acknowledgments}\label{sec:ack}
The authors acknowledge the computational resources provided by the Aalto Science-IT project. Zhiqiang Liao gratefully acknowledges financial support from the Jenny and Antti Wihuri Foundation [grant no. 00230213, 00240203] and the Foundation for Economic Education (Liikesivistysrahasto) [grant no. 230261].

\bibliographystyle{model5-names}\biboptions{authoryear}
\bibliography{References.bib} 

\end{document}